\documentclass[a4paper]{jpconf}

\usepackage{citesort}
\usepackage{graphicx}

\begin{document}
\title{3D Relativistic MHD Simulations of Pulsar Bow Shock Nebulae}

\author{Niccol\`o Bucciantini$^{1,2,3}$, Barbara Olmi$^{1,2,3}$, Luca
  Del Zanna$^{2,1,3}$ }

\address{$^1$ INAF - Osservatorio Astrofisico di Arcetri, Largo E. Fermi 5,
  I-50125 Firenze, Italy\\
$^2$ Dipartimento di Fisica e Astronomia, Universit\`a degli Studi di
Firenze, Via G. Sansone 1, I-50019 Sesto F. no (Firenze), Italy\\
$^3$ INFN - Sezione di Firenze, Via G. Sansone 1, I-50019 Sesto F. no (Firenze), Italy}

\ead{niccolo@arcetri.inaf.it}

\begin{abstract}
Pulsars out of their parent SNR directly interact with the ISM
producing so called Bow-Shock Pulsar Wind Nebulae, the relativistic
equivalents of the heliosphere/heliotail system. These have been
directly observed from Radio to X-ray, and are found also associated
to TeV halos, with a large variety of morphologies. They offer a
unique environment where the pulsar wind can be studied by modelling
its interaction with the surrounding ambient medium, in a fashion that
is different/complementary from the canonical Plerions. These systems
have also been suggested as the possible origin of the positron excess
detected by AMS and PAMELA, in contrast to dark matter. I will present
results from 3D Relativistic MHD simulations of such nebulae. On top
of these simulations we computed the expected emission signatures, the properties of high energy particle escape, the role of current sheets in channeling cosmic rays, the level of turbulence and magnetic amplification, and how they depend on the wind structure and magnetisation.
\end{abstract}

\section{Introduction}

Pulsar Wind Nebulae (PWNe) are synchrotron emitting sources powered by
the wind of a pulsar (PSR). Usually, they are observed inside the
supernova remnant (SNR) of their
parent progenitor, but for old pulsars they can also form as a
consequence of  the direct interaction of the pulsar wind with the
interstellar medium
(ISM) \cite{Gaensler_Slane06a,Bucciantini08b,Olmi:2016}. Pulsar winds
are ultra-relativistic outflows, with typical Lorentz factors in the
range $10^{4}-10^{7}$, 
magnetised, and cold \cite{Goldreich:1969,Kennel:1984,Kennel:1984a}. They are supposed to be
mainly composed by electron-positron pairs
\cite{Ruderman_Sutherland75a,Arons_Scharlemann79a,Contopoulos_Kazanas+99a,Spitkovsky06a,Tchekhovskoy_Philippov+16a,Hibschman:2001,Takata_Wang+10a,Timokhin_Arons13a,Takata_Ng+16a}. 
The interaction
with the ambient medium, forces these supersonic winds to slow
down in a a strong termination shock (TS). It is there that particles are
likely accelerated to a non-thermal distribution \cite{Spitkovsky08a,Sironi:2009,Sironi:2009a,Sironi:2013}. The observed non-thermal
radiation is produced via synchrotron and
inverse Compton scattering, arising from the interaction of these
particles with the magnetic field in the nebula and with the background photon
field.

Given that between 10\% and
50\% of all the pulsars are born with  kick-velocity of the order of $100-500$ km
s$^{-1}$ \cite{Cordes_Chernoff98a,Arzoumanian:2002,Sartore_Ripamonti+10a,Verbunt_Igoshev+17a}, while the supernova remnant expansion is
decelerated \cite{Truelove_McKee99a,Cioffi_McKee+88a,Leahy_Green+14a,Sanchez-Cruces_Rosado+18a}, they are fated to escape the supernova remnant shell
on timescales of the order of a few tens of thousands of years. At
this point their associated nebulae acquire a
cometary-like shape due to the ram pressure balance between the
pulsar wind and the surrounding incoming ISM (in the reference frame of
the PSR) \cite{Wilkin:1996,Bucciantini:2001,Bucciantini:2002}. The pulsar is now
located in the head of these nebulae and an elongated tail forms that extends in the direction
opposite to the pulsar motion. These objects are known as bow
shock PWNe (BSPWNe). The formation of these systems was confirmed by numerical
simulations in different regimes \cite{Bucciantini:2002,Bucciantini:2005,Vigelius:2007,Barkov:2019}.

In the last years BSPWNe have been observed at many different wavelengths,
with a large, and sometimes unexpected, variety of structures at  different
scales: different shapes
and elongation of the tails, different morphologies in the bow shock
head, different polarisation properties, hard X-ray outflows misaligned with
the pulsar velocity \cite{Arzoumanian_Cordes+04a,Gaensler:2004,Yusef-Zadeh:2005,Li:2005,Gaensler05a,Chatterjee:2005,Kargaltsev_Misanovic+08a,Misanovic_Pavlov+08a,Ng_Camilo+09a,Hales_Gaensler+09a,Ng_Gaensler+10a,De-Luca_Marelli+11a,Ng:2012,Marelli_De-Luca+13a,Jakobsen_Tomsick+14a,Auchettl:2015,Klingler_Rangelov+16a,Posselt_Pavlov+17a,Chevalier:1980,Kulkarni_Hester88a,Cordes:93,Bell_Bailes+95a,van-Kerkwijk_Kulkarni01a,Jones_Stappers+02a,Brownsberger:2014,Romani_Slane+17a,Rangelov_Pavlov+16a,Wang_Kaplan+13a}. Interestingly extended TeV halos have also been detected around
BSPWNe \cite{Abeysekara:2017}. Given that these
nebulae could be one of the major contributors of leptonic
anti-matter in the Galaxy, in competition with possible dark matter
sources \cite{Blasi:2011,Amato:2017}, understanding the escape of particles from these
systems, could have important consequences. Contamination from neutral
hydrogen can also modify the dynamics and morphology of the bow-shock
tail, as predicted by theoretical model and verified numerically \cite{Bucciantini:2002a,Morlino:2015,Olmi:2018}.

Beginning with  analytical and semi-analytical
works \cite{Wilkin:1996,Bandiera93a,Bucciantini:2001}, dating  back to more than one
decade, the first numerical models have been presented in the
non-relativistic hydrodynamical regime by \cite{Bucciantini:2002,van-der-Swaluw:2003}, and in relativistic hydrodynamics and magneto
hydrodynamics with spin-kick alinement by
\cite{Bucciantini:2005}. However there is no reason to assume
spin-kick alignment,
and to cope with the relative inclination one needs to work in full 3D. The third dimension is
particularly important to correctly capture the structure of the magnetic
field configuration \cite{Bucciantini:2017}, and  in the study
of the development of turbulence, which can be strongly affected by
geometric constraints. The first 3D models of BSPWNe were presented by \cite{Vigelius:2007} but limited to the classical HD
regime. Only recently 3D MHD simulations have been presented in the
correct relativistic regime \cite{Barkov:2019,Barkov:2019a,Olmi:2019,Olmi:2019a}. More recently \cite{Bucciantini:2018,Bucciantini:2018a},  using a
simplified, axisymmetric laminar semi-analytic model tuned on
numerical simulations, have investigated for then first time how
different magnetic field geometry act on the observed non-thermal
synchrotron emission and polarisation, and on the propagation of
high energy particles, showing the role played by current sheets and current layers.

Here we present the result of a detailed study of the dynamics,
emission, and particle propagation properties of BSPWNe, done using
3D relativistic MHD simulations \cite{Bucciantini:2018,Bucciantini:2018a,Olmi:2019,Olmi:2019a,Olmi:2019b}. Various configurations were
considered in terms of  inclinations of the magnetic field with respect
to the pulsar spin-axis, wind magnetisation, pulsar wind energy
distribution, particle acceleration properties, in order to have a
sample as complete as possible of the interaction conditions.

\section{Numerical Setup and Simulations}

Our simulations were done with the numerical code PLUTO
\cite{Mignone:2007,Mignone:2013}. PLUTO is a shock-capturing, finite-volume code
for hyperbolic and parabolic partial differential equations.
Simulations were done with adaptive mesh refinement (AMR), corresponding to an effective resolution of $2048^3$ cells at the highest level,
sufficient to capture simultaneously the pulsar wind
injection region and the large scale of the cometary tail of the
nebula. A second order Runge-Kutta time integrator and an HLLD Riemann
solver (the Harten-Lax-van Leer for discontinuities,
\cite{Miyoshi:2005,Mignone:2006}) have been used, in order to better
treat shear layers in the nebula. The relativistic pulsar wind is cold
$p/\rho c^2\approx 0.01$, and has a
Lorentz factor $\gamma=10$, high enough to ensure the correct 
relativistic regime. The outer medium is modelled as a cold incoming
flow both with and without magnetic field. Solutions were computed for
various relative inclinations, magnetisations $\sigma$, and pulsar wind energy
distributions, making sure that the dynamics was relaxed to a
quasi-steady regime. For a more detailed description of the setup and
of the models we refer the reader to
\cite{Olmi:2019,Olmi:2019a,Olmi:2019b}, where they are discussed and
presented in more details.  Please nota that the density and magnetic
field can be scaled arbitrarily as far as the ratio, $B^2/\rho c^2$,
is fixed.

The typical scale-length of these nebulae is the so called is the  stand-off
distance, where the wind ram pressure equilibrates the ISM ram
pressure:
\begin{equation}
d_o=\sqrt{\frac{\dot{E}}{4\pi c\rho_{\rm ISM}v_{\rm PSR}^2}}
\end{equation}
where $\dot{E}$ is the pulsar spin-down luminosity, taken equal to the
pulsar wind power, $\rho_{\rm ISM}$ is the ISM density, $v_{\rm PSR}$
the speed of the pulsar with respect to the local medium.

To summarise the main model parameters are:
\begin{itemize}
\item The pulsar is rest at the origin in the Cartesian coordinates,
  while the ISM has the velocity, $(0, 0,-v_{\rm PSR})= (0, 0, -0.1
  c)$, along the $z$-axis.
\item The pulsar spin axis lies in the $y-z$ plane and is offset by
  $\phi_M$ (the spin-kick inclination) from the $z$-axis.
\item The pulsar wind is injected steadily from $r = 0.2 d_o$ either
  with an
  isotropic energy distribution (cases $I$) or following the split-monopole
prescription (cases $A$).
\item The magnetic field injected with the pulsar wind and has only
  toroidal component with respect to the spin axis, while the one in
  the ISM lies in the $y-z$ plane.
\end{itemize}

The magnetisation is defined as $\sigma = B_o^2/(4\pi \rho_o \gamma^2
c^2)$ where, $B_o$ is the strength of the magnetic field in the
equatorial plane of the wind at a distance $r = 0.2
d_o$ (for a split monopole solution), and  $\rho_o$ is the wind
density at the same location \cite{Olmi-Del_Zanna+15a}.

In computing emission maps, the 3D structure of the velocity and magnetic field is provided by our
3D relativistic MHD models. We assume, following \cite{Del-Zanna:2006}, that the emitting
 pairs are distributed according to a power-law  in energy, given the typical high flow speed found in numerical
simulations, even X-ray emitting particles are only marginally
affected by cooling \cite{Bucciantini:2005}, such that our results can reasonably
apply even to higher energies. For simplicity we assume that the power-law index 
is uniform in the nebula. For the emitting particle density we adopt
two different choices: either a uniform distribution, as was done in
\cite{Bucciantini:2018a}, or a density proportional to the local
value of the thermal pressure, as it is customary in other PWNe
models. 
In a few cases we have also investigated a third possibility
that the emission is concentrated in the current sheets that form in
the BSPWN. We build all of our maps using a spectral index
$\alpha=0$. This was shown by \cite{Bucciantini:2018a}
to be a good average for the observed radio spectra, and
changes in the typical observed range do not affect much the results.
The polarised fraction is always given in terms of the theoretical
maximum, that  is $0.7$ for $\alpha=0$. Emission maps where computed for all simulations. Emissivity toward the observer is computed taking into
account all relativistic effects, including Doppler boosting and
aberration. For a complete description of the method used and the
formalism adopted we refer to \cite{Del-Zanna:2006,Volpi:2007,Bucciantini:2005a,Olmi:2014}. We have also computed the
polarisation properties, accounting for relativistic polarisation angle swing.

In computing particles escape, given that we are interested in the escape  from
the very head of these nebulae, as in \cite{Bucciantini:2018a}, particles that move
backward at a distance exceeding $10-20d_o$,
from the PSR are assumed to be lost in the tail. It is still likely that those particles
can escape from the tail, but given that the tail might be in general
more turbulent than the head, we expect the escape there to be more
isotropic. Moreover, we assume that all high energy particles
are injected from the PSR (or equivalently the pulsar wind) in the
radial direction. The electric field is given by the ideal MHD
condition ${\bf E}=-{\bf V}\times {\bf B}/c$ where
{\bf V} and {\bf B}  are the flow speed and magnetic
field given by our RMHD simulations. All particles are injected with
the same Lorentz factor, and we have investigated four cases with
$\gamma=[0.5,1.0,3.0,10.]\times 10^7$, corresponding to values where
we expect the transition from Larmor radii in the magnetic field
smaller than the size of the bow-shock set by its stand-off
distance $d_o$, to larger than $d_o$. Particle trajectories in the electric and magnetic fields of this system are computed using an explicit Boris Pushing technique \cite{Boris:1972,Vay:2008,Higuera:2017}, which ensures energy and phase space number conservation in the non-radiative regime. We verified, by computing radiative losses, that they are always negligible.

\section{Dynamics}

We summarise here the results of our numerical study on the dynamics,
morphology, and magnetic field structure in BSPWNe. More details can
be found in \cite{Olmi:2019}.

\begin{figure}[h]
  \includegraphics[width=.90\textwidth,clip,bb=40 150 500
   650]{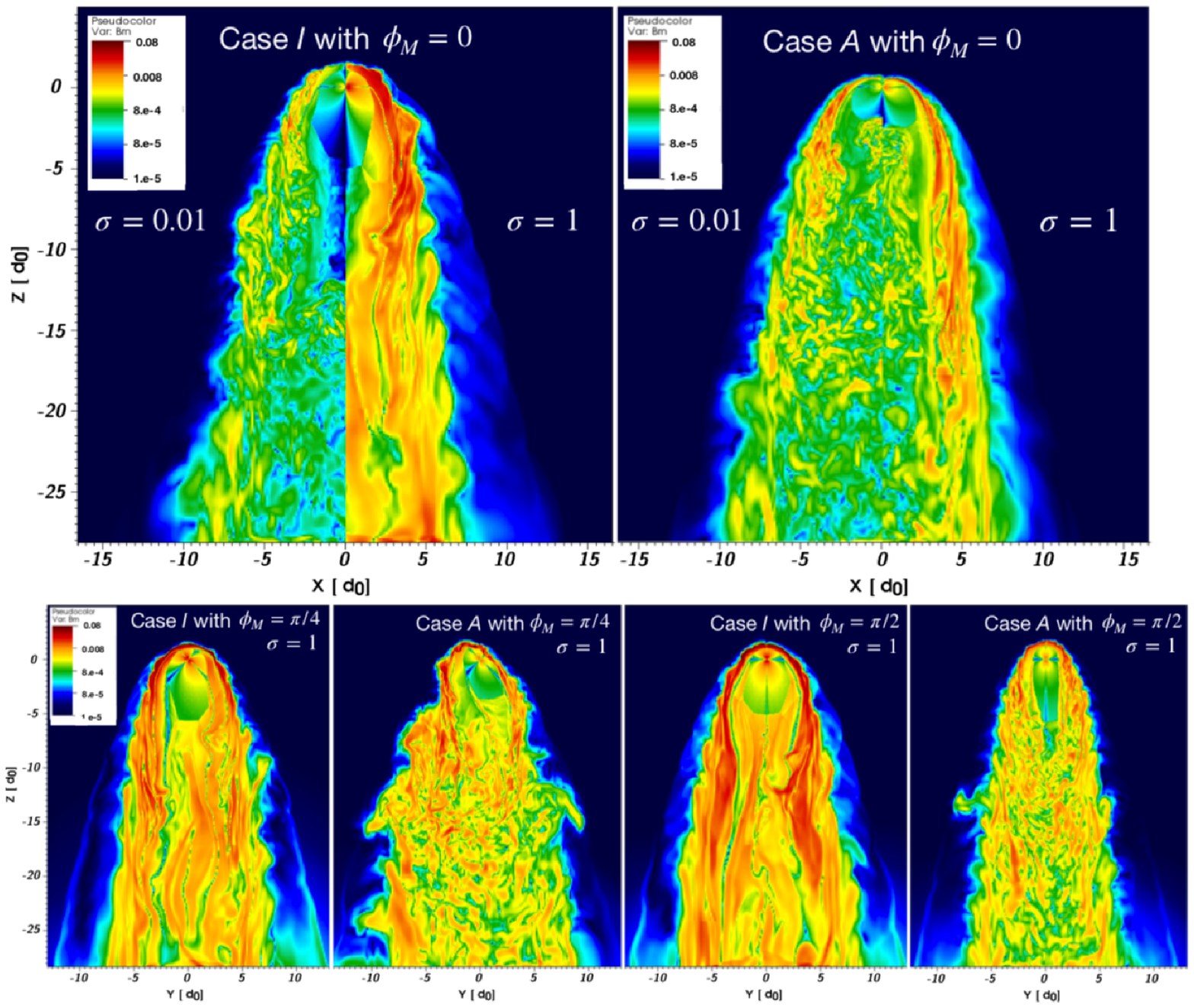}
	\vspace{-0.3cm}
\begin{minipage}[b]{35pc}\caption{\label{Fig:1}
Map of the magnetic field strength in the $y=0$ plane (in code units), normalised to the
maximal nominal value. Upper panels: comparison for runs $I$
(on the left) and $A$ (on the right), for  in cases where the pulsar
spin-axis is aligned to its kick velocity; maps are composed by
two halves of different value of magnetisation, lower on the
left-side and higher on the right-side, respectively. Bottom panels:
maps of the magnetic field strength for magnetisation $\sigma=1.0$ and
for the all the remaining inclinations $\phi_M$, both for $I$ and $A$ cases. From \cite{Olmi:2019}.}
\end{minipage}
\end{figure}

Comparing 3D with 2D HD runs we find a good agreement for isotropic
winds, especially at intermediate values of the magnetisation ($\sigma=0.1$). 
This is unexpected, since in principle one would expect a larger similarity in the low magnetisation cases.
However for values of the magnetisation $\sigma<0.1$ the dynamics is
completely dominated by turbulence on small scales. On the other hand
for larger values,
 the conditions at injection become dominant.
In the tail it is quite clear the difference in the development of the local turbulence and in the
level of mixing of the fluid, as shown in Fig.~\ref{Fig:1}.

In the direction aligned with the pulsar motion ($z-$direction) the
velocity structure shows a lower velocity channel around the $z$ axis
($\sim 0.65c$ in 2D and $\sim 0.7-0.8c$ in 3D) surrounded by an higher
velocity flow layer (with $\sim 0.85c$ in 2D and up to $0.9c$ in
3D). In the $A$ cases the effect of turbulence is even more
pronounced: injection matters only for $\sigma>0.1$, and the
typical structures in the tail remain coherent on shorter scales than
in the corresponding $I$ case. There is also a higher level of
mixing between the pulsar wind and ISM material.

In the $I$ cases the initial magnetic field configuration, and
the current sheets separating  field of opposite polarities, survive
in the tail, while in the $A$ ones turbulent mixing tends to
destroy them even at higher magnetisation.
We found that there is no efficient turbulent magnetic field
amplification: in the presence of turbulence it tends just to reach
equipartition with the turbulent kinetic energy, which is usually
smaller than the thermal energy in the tail.
In 3D there is evidence of magnetic field enhancement near to the contact
discontinuity (CD), possibly the effect of an efficient shear instability
amplification acting at the CD.

The forward shock shape is almost the same, the
only exception being the $A$ cases with the spin-kick
inclination $\phi_M=\pi/4$:
differences appear as small extrusions and blobs, possibly resulting
as periodic perturbation of the forward shock. Fluctuations of the flow in the
interior are not characterised by high values of the magnetic field, density or pressure.
Moreover they arise on very small spatial scales,
possibly making it difficult to be revealed by actual instruments due
to resolution limits. 

Major differences among various cases  are visible mainly in terms of collimation or broadness
of the tails, arising as the effect of different magnetic inclination
and especially of the wind anisotropy. 
The average flow pattern in the tail  between 2D and 3D runs is not
significantly different (differences are largely in the head due to inclination and
anisotropy). In the low $\sigma$ regime there is a strongly turbulent
magnetic field structure, suggesting that perhaps a laminar model is
not likely to fully capture the magnetic field. For higher values of
the magnetisation $\sigma$, the magnetic field is more coherent, qualitatively in agreement with the prediction of simplified laminar models \cite{Bucciantini:2018}.

\section{Emission}

We summarise here the results of our numerical study on the emission
and polarisation properties of BSPWNe. More details can
be found in \cite{Olmi:2019}.

Using our numerical simulations we analysed both cases of isotropic  ($I$) and
anisotropic ($A$) winds. In the case of uniform injection and high magnetisation,
we found a strong correlation between the conditions at injection,
as the inclination $\phi_M$, and the surface brightness of our simulated
maps. The trend in general agrees with that of fully laminar
semi-analytic models \cite{Bucciantini:2018}. There is a
variety of morphologies, from systems brighter in the head to ones
dominated by the tail. Some maps show
bright wings. In computing the maps one needs also to consider the
viewing angle $\chi$, however the dependence on it of the observed properties is less marked
and appreciable only at high resolution. In the high $\sigma$ 
regime the polarised fraction is quite high in the
tail.  Once the magnetisation drops, turbulence 
begins to dominate the appearance of the maps, and it is far more difficult to
find defined observational patterns.  In this cases it will be hard to
distinguish between different inclinations $\phi_M$ in a robust way. 

\begin{figure}[h]
\centering
	\includegraphics[width=.90\textwidth,clip,bb=40 150 700
   650]{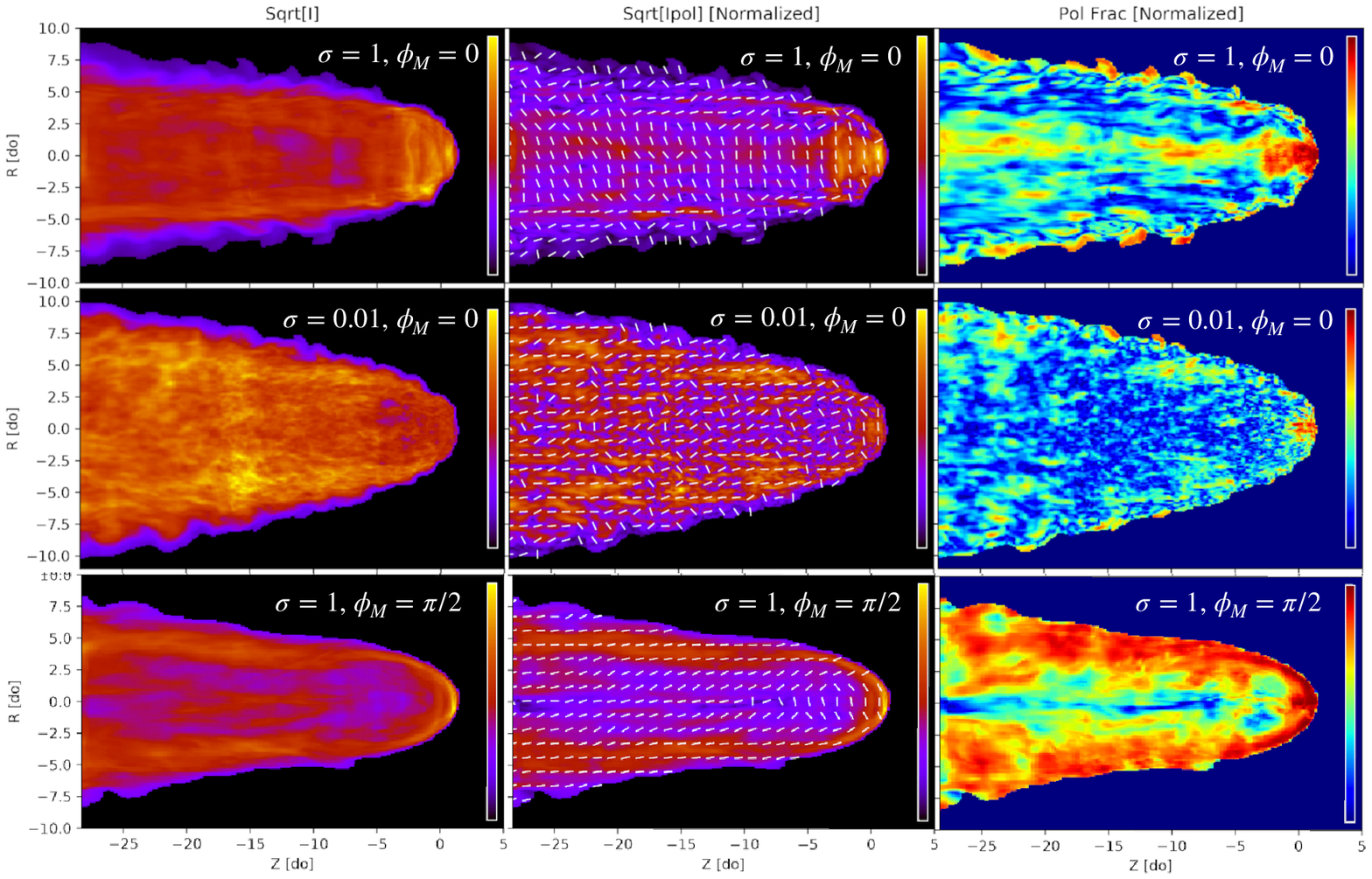}
        	\vspace{-0.3cm}
\begin{minipage}[b]{35pc}\caption{\label{Fig:2}	
Emission maps for the case of a uniform wind with pulsar spin-axis aligned with
the kick velocity and high magnetisation (upper row); the
same but for low magnetisation (middle row), and
finally for a pulsar spin-axis orthogonal  to the pulsar the kick
velocity again at high magnetisation (bottom row; viewed
from the pulsar axis direction). All cases are computed assuming
uniform local emissivity. From left to right: square root of the total
synchrotron intensity normalised to the maximum, square root of the
polarised intensity normalised to the maximum superimposed with the
polarised direction, and polarised fraction normalised to the
theoretical maximum for a power-law synchrotron with $\alpha = 0$. The
colour scale is linear between zero and one. From \cite{Olmi:2019a}. }
\end{minipage}
\end{figure}

Interestingly in our maps, apart from the strongly
turbulent cases, we do not observe major time variations
along the tail direction. This implies that time variability
and temporal changes in the flow pattern are weak, and unlikely to give
rise to major observational changes. On the other hand,
changes in the polarisation properties of a BSPWN could likely point
to a strongly anisotropic energy injection, and spin
axis misalignment.

Our models were computed according to different scalings for the
emitting particles density. If we consider a scaling proportional
to the local pressure, typical of
particles accelerated at the wind termination shock and then
advected in the nebula, we find that the head is much brighter than the
tail, even by a factor 10. Only in systems dominated by turbulence this
difference is less enhanced. Indeed system like the Mouse
nebula show a very bright head and a fainter tail, but in many others
there is no evidence for such brightness difference \cite{Yusef-Zadeh:2005,Ng:2012}. This
could be the signature of an  acceleration process
diffused in the bulk of the nebula. We also tried an emissivity scaled
according to the strength of the currents, aiming at modelling the
effect of reconnection and dissipation of currents. However there seem to be no appreciable
difference with respect to  a uniform emissivity.

In almost
all cases the direction of the polarisation (the inferred direction of
the magnetic field) seems to be almost aligned with the tail. This is
an interesting aspect, 
most likely due to polarisation swing associated to the relativistic
flow in the tail. If one suppresses relativistic beaming and
aberration, when computing maps, the structure of the polarisation
pattern changes, and the polarisation looks less aligned. Changes in the polarisation direction are
observed in many BSPWNe, and they are usually associated with changes in brightness \cite{Ng_Gaensler+10a}. Our
results suggest that these could originate when the flow decelerates
maybe as a consequence of internal shocks, or because of mass loading
from the ambient medium. 

\section{Particle Escape}

 In few cases, non-thermal high energy emission is observed
 outside of the supposed location of the contact discontinuity
 separating the pulsar material from the ISM, where the canonical
 models predict there should be none. This emission ranges from faint
 X-ray haloes, as in the case of IC443 \cite{Swartz_Pavlov+15a} and
 the Mouse \cite{Gaensler_van-der-Swaluw+04a}, or large non-thermal
 TeV haloes like in the case of Geminga
 \cite{Posselt_Pavlov+17a,Abeysekara_Albert+17a}, to more structured
 features like the X-ray {\it prongs} observed ahead of the bow shock
 in G327.1-1.1 \cite{Temim_Slane+15a}, or the one sided jets as seen
 in the Guitar Nebula \cite{Hui_Becker07a} or in the Lighthouse
 nebula  \cite{Pavan_Bordas+14a}.

\begin{figure}[h]
\centering
\includegraphics[width=.90\textwidth,clip,bb=40 150 700
   650]{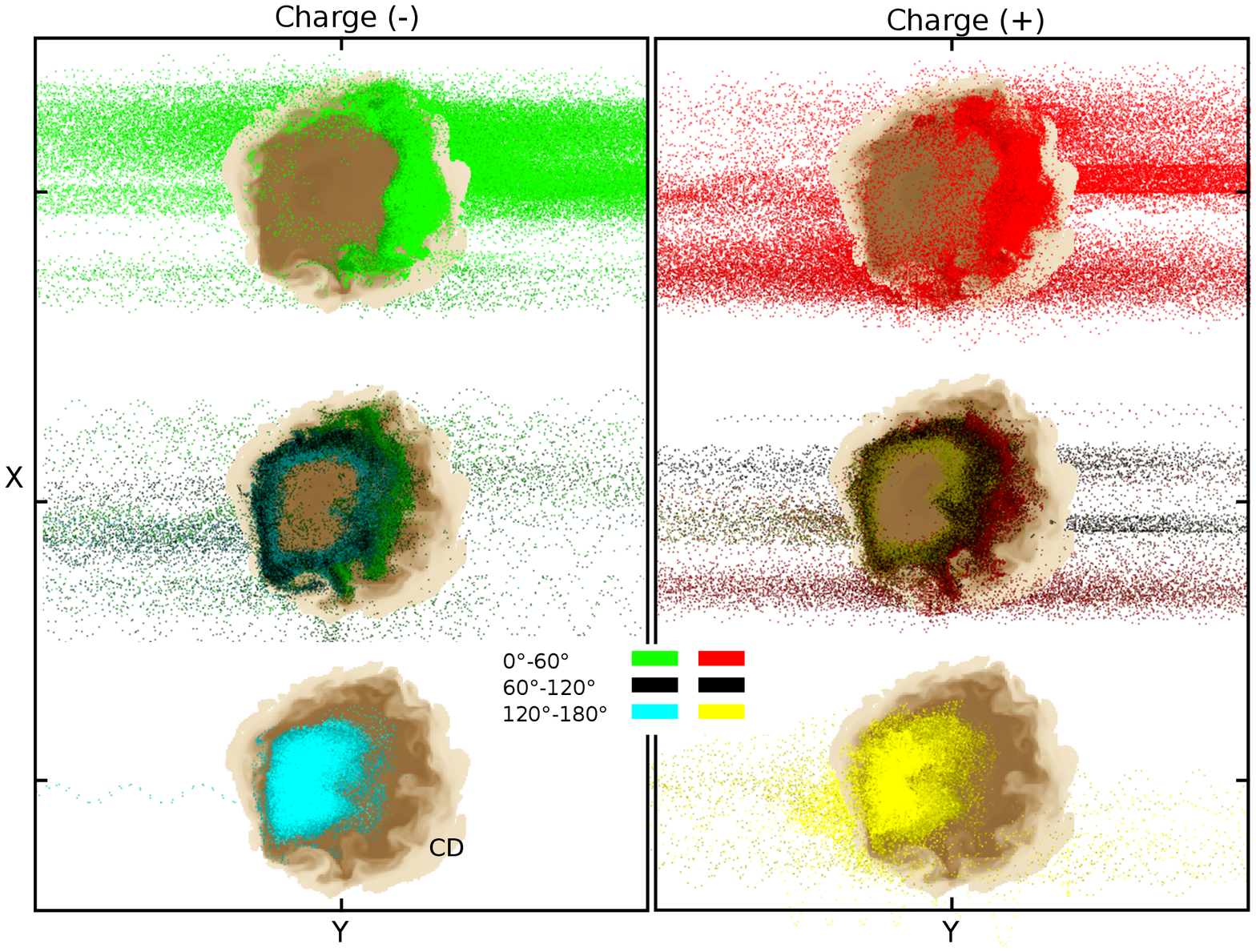}
            	\vspace{-0.3cm}
        \begin{minipage}[b]{35pc}\caption{\label{Fig:3}
Projection on the plane transverse to the pulsar motion of the 3D-positions of particles
injected in the wind with a Lorenz factor $\gamma=3\times
10^7$. Left and right panels refers to particles with different
signs. From top to bottom, particles injected within different ranges
of the polar angle $\theta$ with respect to the pulsar spin-axis:
$[0,60]^\circ$ (green and red) indicates particles injected along the
polar current originating from the PSR spin-axis pointing toward the
PSR motion,   $[60,120]^\circ$ (black)  indicates particles injected
along the equatorial current sheet,   $[120,180]^\circ$ (cyan  and
yellow)  indicates particles injected along the polar current pointing
in the opposite direction with respect to the PSR motion. The
background image shows the $\log_{10}$  cut of the density (darker brown
for lower values, lighter brown for higher ones), in the tail at a
distance from the pulsar  $z=-11.5 d_0$
within the position of the contact discontinuity with the ISM material, in order to mark the location of the
shocked pulsar wind. From \cite{Olmi:2019b}.}
\end{minipage}
\end{figure}

Here we illustrate the results obtained by computing the trajectories of
high-energy particles in the electric and magnetic field of our
simulations, to investigate their possible escape.  Results are shown in Fig.~\ref{Fig:3}. The typical energy scale of
 pairs is set by the condition that their Larmor radius in the
 equipartition magnetic field in the head is equal to the typical size
 of the bow-shock \cite{Bandiera08a}, and this 
 corresponds, for typical system, to a Lorentz factor $\simeq 3
 \times 10^7$. 

 Our models show a transition in the properties of escaping particles,
 most of whom comes from the frontal polar region
 of the pulsar wind, while the
 others tend to remain confined in the tail. At low energies the escape
 looks more likely due to reconnection of magnetic field lines between the
 pulsar wind and ISM: particles moving along those
 magnetic field lines can escape into the ISM. The outflows
 are asymmetric (by a factor 4 to 5), likely a consequence of the reconnection at the magnetopause.

At higher energies the escape enters a different regime. 
It becomes more charge separated. Now current sheets and layers play a
more important role, as suggested in  \cite{Bucciantini:2018a}. 
The charge asymmetry is about a factor 1.5, while the spatial
asymmetry is a factor 2. 

At very high energies the regime becomes fully diffusive
\cite{Bykov:2017}: the charge asymmetry exceeds a factor 2, while the
spatial asymmetry is reduced to less than 1.5. This transition appears
to take place within just an order of magnitude in the energy of the particles
from $\gamma=10^7$ to $\gamma=10^8$, suggesting that their outflow is 
quasi-monochromatic. 

Both the asymmetry in the escaping flux and it being charge separated, can explain the presence of one-sided jet.
Unfortunately the relation between the number of escaping
particles and the presence of bright features is non trivial: magnetic
field amplification is required, together with turbulent particles
confinement, in a non-linear regime
\cite{Bandiera08a}. This complex physics goes beyond what we can
simulate at the moment. For example, in the presence of a net current, the magnetic field can be
amplified more efficiently in the non-resonant regime \cite{Skilling71a,Bell04a}. 
This means that self confinement is more efficient. 

Our results also
show what could be a possible explanation for the X-ray morphology of
G327.1-1.1 \cite{Temim_Slane+15a}, and the two long tails seen in
Geminga \cite{Posselt_Pavlov+17a} (not due to limb brightening).

\section{Conclusion}
We present here the summary of a detailed numerical study of bow-shock
pulsar wind nebulae, carried in 3D using relativistic MHD. Our study
was focused on the development of a complete set of models in terms
of injection properties, and ISM conditions, that allowed us to
investigate not only how the dynamics changes in response to
different injections (in terms of spin-axis inclinations,
magnetisation, anisotropy)  but also to compute on top of these fluid
models realistic emission maps, using different prescriptions for the
distribution of emitting particles, and to evaluate also the escape
properties of high energy pairs, and the possible origin of extended
non-thermal features.

Our results show that the large variety of observed morphologies, and
structures in known pulsar bow-shocks, can be reasonably well accounted
for in terms of differences in the injection conditions. We plan in the
future to continue our investigation of few selected configurations,
that looks more representative of specific known objects.

\subsection{Acknowledgments}
We acknowledge the ``Accordo Quadro INAF-CINECA (2017-2019)''  for  high performance computing resources and support. Simulations have been performed as part of the class-A project ``Three-dimensional relativistic simulations of bow shock nebulae'' (PI B. Olmi). 
The authors acknowledge financial support from the ``Accordo Attuativo ASI-INAF n. 2017-14-H.0 Progetto: \textit{on the escape of cosmic rays and their impact on the background plasma}'' and from the INFN Teongrav collaboration.

\section*{References}
\bibliography{olmi}

\end{document}